\documentclass[conference,onecolumn]{IEEEtran}
\usepackage{cite}
\usepackage{amsmath,amssymb,amsfonts}
\usepackage{algorithmic}
\usepackage{graphicx}
\usepackage{textcomp}
\usepackage{xcolor}
\usepackage{listings}
\usepackage{xspace}
\usepackage[colorlinks=false,]{hyperref}
\usepackage{placeins}
\usepackage{tikz}

\usepackage[printwatermark]{xwatermark}
\newwatermark[allpages,color=gray!50,angle=45,scale=2,xpos=0,ypos=0]{Preprint}

\newboolean{showcomments}
\setboolean{showcomments}{true}		
\ifthenelse{\boolean{showcomments}}
{\newcommand{\nb}[2]{
		\fbox{\bfseries\sffamily\scriptsize#1}
		{\sf\small$\blacktriangleright$\textcolor{red}{#2}$\blacktriangleleft$}
	}
}
{\newcommand{\nb}[2]{}
	
}
\usepackage[normalem]{ulem}
\newcommand{\rst}{Restats\xspace}

\newcommand{\openapi}{OpenAPI\xspace}

\newcommand{\twoxx}{\texttt{2XX}\xspace}

\newcommand{\fourxx}{\texttt{4XX}\xspace}

\newcommand{\fivexx}{\texttt{5XX}\xspace}

\newcommand{\eg}{e.g.,~}
\newcommand{\ie}{i.e.,\ }

\renewcommand{\descriptionlabel}[1]{\hspace{\labelsep}\textit{#1}}

\usepackage{pifont}

\definecolor{delim}{RGB}{20,105,176}
\definecolor{numb}{RGB}{0, 51, 0}
\definecolor{string}{RGB}{77, 0, 77}

\lstdefinelanguage{json}{
	numbers=left,
	numberstyle=\small,
	frame=single,
	rulecolor=\color{black},
	showspaces=false,
	showtabs=false,
	breaklines=true,
	postbreak=\raisebox{0ex}[0ex][0ex]{\ensuremath{\color{gray}\hookrightarrow\space}},
	breakatwhitespace=true,
	showstringspaces=false,
	basicstyle=\ttfamily\footnotesize,
	upquote=true,
	morestring=[b]",
	stringstyle=\color{string},
	literate=
	*{0}{{{\color{numb}0}}}{1}
	{1}{{{\color{numb}1}}}{1}
	{2}{{{\color{numb}2}}}{1}
	{3}{{{\color{numb}3}}}{1}
	{4}{{{\color{numb}4}}}{1}
	{5}{{{\color{numb}5}}}{1}
	{6}{{{\color{numb}6}}}{1}
	{7}{{{\color{numb}7}}}{1}
	{8}{{{\color{numb}8}}}{1}
	{9}{{{\color{numb}9}}}{1}
	{\{}{{{\color{delim}{\{}}}}{1}
	{\}}{{{\color{delim}{\}}}}}{1}
	{[}{{{\color{delim}{[}}}}{1}
	{]}{{{\color{delim}{]}}}}{1},
}

\lstdefinelanguage{http}{
	numbers=left,
	numberstyle=\small,
	frame=single,
	rulecolor=\color{black},
	showspaces=false,
	showtabs=false,
	breaklines=true,
	postbreak=\raisebox{0ex}[0ex][0ex]{\ensuremath{\color{gray}\hookrightarrow\space}},
	breakatwhitespace=true,
	showstringspaces=false,
	basicstyle=\ttfamily\footnotesize,
	upquote=true,
	morestring=[b]",
	stringstyle=\color{string},
}

\newcommand\YAMLcolonstyle{\color{red}\ttfamily\small}
\newcommand\YAMLkeystyle{\color{black}\ttfamily\small}
\newcommand\YAMLvaluestyle{\color{blue}\ttfamily\small}

\lstdefinelanguage{yaml}{
	keywords={true,false,null,y,n},
	keywordstyle=\color{darkgray},
	basicstyle=\YAMLkeystyle,                                 % assuming a key comes first
	frame=single,
	numbers=left,
	numberstyle=\small,
	sensitive=false,
	comment=[l]{\#},
	morecomment=[s]{/*}{*/},
	commentstyle=\color{purple},
	stringstyle=\YAMLvaluestyle,
	showstringspaces=false,
	moredelim=[l][\color{orange}]{\&},
	moredelim=[l][\color{magenta}]{*},
	moredelim=**[il][\YAMLcolonstyle{:}\YAMLvaluestyle]{:},   % switch to value style at :
	morestring=[b]',
	morestring=[b]",
	literate =    {---}{{\ProcessThreeDashes}}3
	{>}{{\textcolor{red}\textgreater}}1
	{|}{{\textcolor{red}\textbar}}1
	{\ -\ }{{\mdseries\ -\ }}3,
}
\begin{document}

\title{
% \rst: A Measurement Tool to Empirically Compare Automated Test Case Generation Approaches for RESTful APIs
\rst: A Test Coverage Tool for RESTful APIs
%\thanks{Blind funding agency}
}

%\author{\IEEEauthorblockN{Davide Corradini}
%\IEEEauthorblockA{\textit{Department of Computer Science} \\
%\textit{University of Verona}\\
%Verona, Italy \\
%davide.corradini@univr.it}
%\and
%\IEEEauthorblockN{Amedeo Zampieri}
%\IEEEauthorblockA{\textit{Department of Computer Science} \\
%\textit{University of Verona}\\
%Verona, Italy \\
%amedeo.zampieri@studenti.univr.it}
%\and
%\IEEEauthorblockN{Michele Pasqua}
%\IEEEauthorblockA{\textit{Department of Computer Science} \\
%\textit{University of Verona}\\
%Verona, Italy \\
%michele.pasqua@univr.it}
%\and
%\IEEEauthorblockN{Mariano Ceccato	}
%\IEEEauthorblockA{\textit{Department of Computer Science} \\
%\textit{University of Verona}\\
%Verona, Italy \\
%mariano.ceccato@univr.it}
%}
\author{%
\IEEEauthorblockN{%
Davide Corradini\IEEEauthorrefmark{1}, 
Amedeo Zampieri\IEEEauthorrefmark{2}, 
Michele Pasqua\IEEEauthorrefmark{3} and 
Mariano Ceccato\IEEEauthorrefmark{4}}
\IEEEauthorblockA{%
\textit{Department of Computer Science} \\
\textit{University of Verona} -- Verona, Italy \\
Email: %
\IEEEauthorrefmark{1}davide.corradini@univr.it, 
\IEEEauthorrefmark{2}amedeo.zampieri@studenti.univr.it, 
\IEEEauthorrefmark{3}michele.pasqua@univr.it, 
\IEEEauthorrefmark{4}mariano.ceccato@univr.it}
}

\maketitle

\begin{center}
	\begin{tikzpicture}
	\node (A) at (0,1.25) {Paper accepted for publication in the proceedings of:};
	\node at (0,0.75) {\emph{37$^\text{th}$ IEEE International Conference on Software Maintenance and Evolution (ICSME 2021)}};
	\node (B) at (0,0) {\footnotesize The present document is the preliminary version of the work prior to peer-review. The final version can be found on the publisher website.};
	\filldraw[rounded corners=2pt,fill=gray,draw=gray!25,opacity=0.25] (B.south west) rectangle (B.east |- 52, 52 |- A.north);
	\end{tikzpicture}
\end{center}

\bigskip

\begin{abstract}
Test coverage is a standard measure used to evaluate the completeness of a test suite. Coverage is typically computed on source code, by assessing the extent of source code entities (e.g., statements, data dependencies, control dependencies) that are exercised when running test cases. %coverage is probably the most accurate way to achieve this objective, an alternative where a multitude of languages, frameworks and micro-services are usually deployed, it is often difficult to adopt this metric.
When considering REST APIs, an alternative perspective to assess test suite completeness is with respect to the service definition.%, since accessing REST APIs source code is usually not possible.

This paper presents \rst, a test coverage tool for REST APIs that supports eight state-of-the-art test coverage metrics with a black-box perspective, \ie only relying on the OpenAPI interface specification of the REST API under test. In fact, metrics are computed by only observing the HTTP requests and responses occurring at testing time, and no access to source/compiled code of the REST API is required.

These coverage metrics come in handy for: (i) developers and test engineers working at development and maintenance~tasks; (ii) stakeholders and customers who want to evaluate the completeness of acceptance tests; (iii) researches interested in comparing different automated test case generation strategies. %As a matter of fact, these coverage metrics have been adopted in our previous work of comparing different automated test case generation tools for REST APIs.

\rst GitHub repository: \url{https://github.com/SeUniVr/restats}

\rst demo video: \url{https://smarturl.it/restats-demo}

\end{abstract}


\section{Introduction}
\label{sec:intro}

% \mariano{Structure: \\ 
% - description of the context, positioning \\
% - motivation, highlighting the relevance to software evolution, i.e. automated testing\\
% - usage scenario \\
% - novelty and relation to previous research}

A RESTful API (or REST API for short) is an API that respects the REST (REpresentational State Transfer) architectural style~\cite{fielding2000architectural}. REST APIs provide a uniform interface to create, read, update and delete (CRUD) a resource. A resource is generally identified by an HTTP URI, and CRUD operations are usually mapped to the HTTP methods POST, GET, PUT and DELETE to the resource URI.

%The resource URI and the HTTP methods may accept input parameters, to specify additional information for executing the API operations, such as the identifier of the object to retrieve or a structured object to be added to the collection using the POST method.

REST APIs are becoming a de-facto industrial standard to interconnect different computer systems. They are used in a quite wide set of contexts, \eg when exchanging data with the cloud~\cite{APIcloud2016}, when connecting smartphone apps to their corresponding server~\cite{APImobile2019}, for identity provisioning~\cite{SCIM} and when inter-operating different banks~\cite{OBP}. 

The correct integration among computer systems, potentially carried out by different parties, is a critical point. To monitor correct integrations and promptly reveal implementation defects, several tools are available for testing computer systems through their REST APIs. These tools include not just extensions to general purpose testing frameworks (\eg REST-Assured~\cite{RESTAssured}, that extends the popular JUnit testing framework with REST API specific features), but also brand-new tools, explicitly developed to test REST APIs (\eg Postman~\cite{Postman} and SoapUI~\cite{Soap}). However, they still require developers to spend valuable time in manually coding test cases. Additionally, to mitigate the cost of manually writing all the test cases, the research community proposed several approaches to automatically generate test cases for REST APIs (\eg QuickREST~\cite{Karlsson2020QuickREST}, RESTler~\cite{Atlidakis2019RESTlerSR} and RestTestGen~\cite{Viglianisi2020RESTTESTGENAB}). %Some have also been implemented into software tools, and each one of these adopts its own, different approach for creating test suites and assessing successful test cases and faulty ones, implementing specific oracles. 

Despite several approaches being available, either to generate and run test cases on REST APIs, to the best of our knowledge no tool is available to {\em measure} test coverage, \ie to quantify the extent of a REST API that has been subject to testing and what part of it still requires more tests. To fill this gap, we propose \rst, a tool meant to monitor the whole REST API testing process and compute the corresponding comprehensive coverage report.

\rst is meant to be helpful in many contexts.
\begin{itemize}
	\item Supporting software developers when developing and/or evolving software systems that expose a REST API. In fact, before publishing an update, a REST API should be carefully assessed and its tests should meet an adequate level of coverage.
	\item Supporting stakeholders and customers in evaluating REST APIs delivered by contractors or by third parties, before fully committing and adopting them. In fact, stakeholders might be interested in assessing the coverage of their acceptance tests before accepting a product.
	\item Supporting researchers in conducting experimental surveys, when comparing different tools and alternative strategies to automatically generate test cases for REST APIs. In fact, an approach that delivers test cases with higher coverage might have potentially higher chances of exposing defects in the implementation of a REST API.
\end{itemize}

Despite theoretical metrics have been proposed to measure REST APIs test coverage~\cite{MartinLopez2019TestCC}, no implementation is available to actually compute them. So, although theoretically sound, in practice REST APIs test coverage metrics convey limited value to developers and practitioners.
%\rst has been developed as part of our research, to allow us to conduct the experimental comparison of automated testing tools~\cite{ICSME2021unpub}. In fact, each testing tool uses its own metric to asses the quality of the testing process, resulting in reports incompatible with each other and very hard to compare. A common coverage tool allowed us to conduct a sound and objective comparison on a common ground. 
To the best of our knowledge, \rst is the first tool that computes coverage metrics for REST API test suites. 

This paper is organized as follows. In Section~\ref{sec:architecture} we present the architecture of \rst and the implemented coverage metrics. In Section~\ref{sec:runningexample} we show how to use \rst on a case study, reporting the output of the tool (\ie the computed coverage metrics) for the REST API under test. Section~\ref{sec:related} presents related work  and, finally, Section~\ref{sec:conclusion} closes the paper.

\section{Architecture}
\label{sec:architecture}

%\mariano{Structure: \\ 
%- overview of architecture\\
%- description of components \& description of inner working
%- summary of available metrics}

\begin{figure}[t]
	\caption{\rst architecture overview.}
	\label{fig:restats_flow}
	\centering
	\includegraphics[width=.65\linewidth]{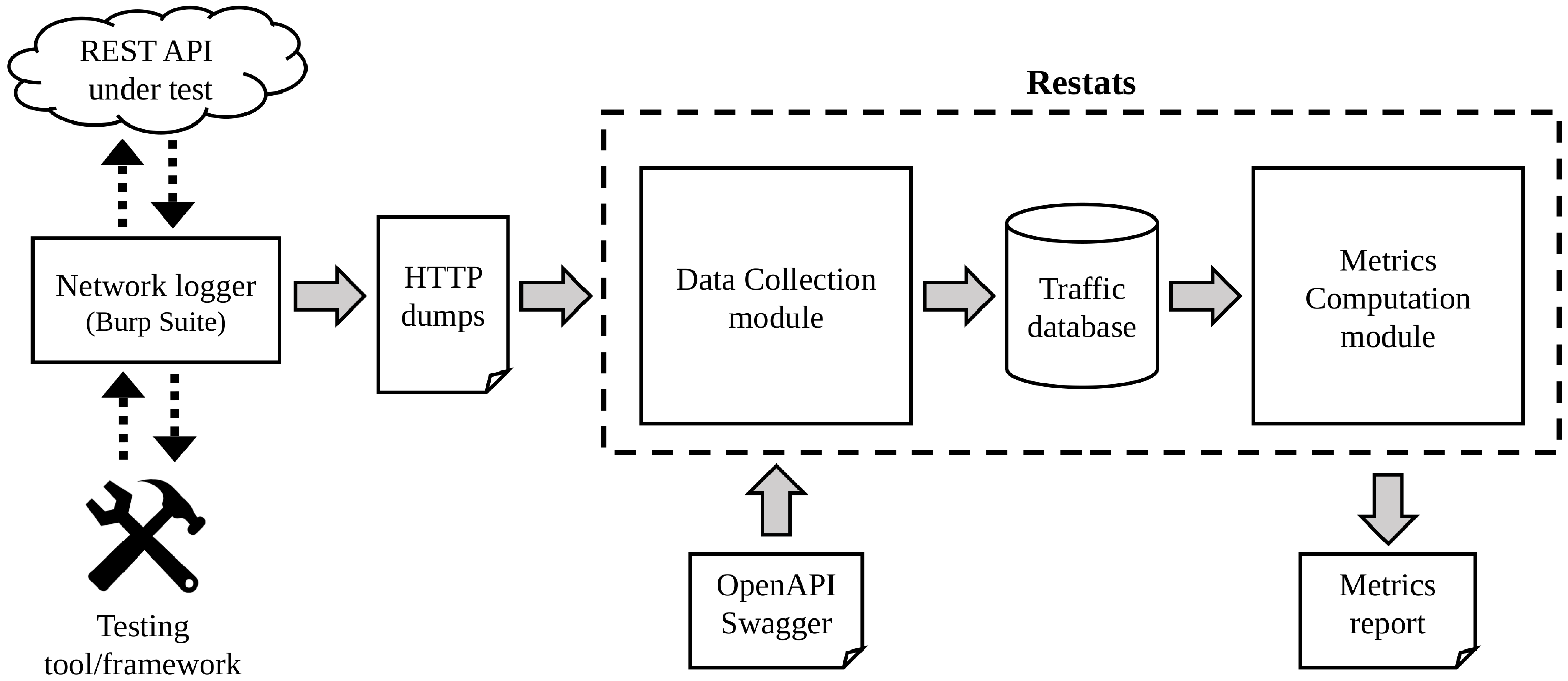}
\end{figure}

\rst computes REST API test coverage metrics based on the theoretical measurement framework proposed by Martin-Lopez et al.~\cite{MartinLopez2019TestCC}. An overview of the architecture of \rst is shown in Figure~\ref{fig:restats_flow}. The tool requires as input the REST API definition file (called OpenAPI specification or Swagger) and a network log containing the HTTP messages exchanged between the API and the testing tool. An external {\em Network logger} is in charge of collecting these messages. In particular, we used the HTTP proxy offered by the Burp Suite~\cite{Burp}.

The network log is consumed by the \emph{Data Collection module}, in which raw inputs are rearranged, processed and written to a database. Eventually, the \emph{Metrics Computation module} computes the coverage metrics using data from the database and fills in a final coverage report.

\subsection{Network Logger}

To collect testing input and output, we use the \emph{Burp Suite}~\cite{Burp}. It works as a proxy and collects all the HTTP messages exchanged between the testing tool and the REST API under test. 
%The proxy features an HTTP history, in which all the parsed traffic is displayed. 
Then, using the \emph{burp-dump}~\cite{BurpExtension} extension, we dump all HTTP requests and HTTP responses as plain text. Note that \rst is not dependent on the Burp Suite; any alternative tool that can emit a plain-text log of HTTP messages is fine. The user can choose an alternative component that best fits his/her needs to gather request/response samples, as long as they can be dumped as text file(s). HTTP messages dumps are expected to follow a naming convention to pair requests and responses in the form $(\text{\emph{n-request.txt}},\text{\emph{n-response.txt}})$, with a progressive identifier \emph{n}.

\subsection{Data Collection Module}
This module is used to parse network dumps. Given the textual HTTP dumps as input, it extracts all the information needed for the coverage computation (\eg request paths, HTTP methods, response status codes). Then, it populates a SQLite %\footnote{We chose SQLite as RDBMS due to its performance, portability and configuration-less design.} 
database. 
%(optionally, a JSON file with the parsed request-response  pairs can be created). 

To this aim, HTTP requests and responses are parsed with the following strategies.
\begin{LaTeXdescription}
	\item[Message payload.] HTTP requests and responses might contain arbitrary content types in their body,  either text (\eg JSON, XML) or binary (\eg JPEG, MP4), leading to possible message body parsing failures. Therefore, message body is safely parsed as a binary content. A second parser is called for the message body, depending on the content type declared in the message header. At the moment, only JSON payloads are supported. Hence, if a different format is found in the header, the payload is ignored.
	
	\item[Unpaired requests.] Not all HTTP requests match a corresponding HTTP response. In fact, some requests could have been ignored due to a too high server load or a server crash. Even if unpaired, these orphan requests are still recorded in the database. Indeed, they are used for the computation of input coverage metrics, which do not require a matching response. %\mariano{non capisco il senso di questo secondo punto...} \davide{Vengono utilizzate per calcolare le input coverage metrics che non hanno bisogno della risposta}
	\item[Path parameter matching.] The \openapi standard allows path templating, \ie they allow to specify input data as part of the called URL. Therefore, \rst has to be able to group and recognize HTTP requests belonging to the same path template. This task is challenging to carry out in the case of \openapi specifications with colliding paths. For instance, \texttt{/users/\{id\}} (where \{id\} is a placeholder for any string) and \texttt{/users/auth} seems referring to the same path, and \texttt{auth} could be interpreted as both a specific path and a user id.
\end{LaTeXdescription}

All the fields of HTTP requests and responses are then saved in the database to be quickly accessed by the subsequent module. We decided to use a SQLite database because of its performance, portability and configuration-less features.

\subsection{Metrics Computation Module}

This module takes the database populated by the Data Collection module and the \openapi specification as input. The database is used to recognize which elements of the REST API have been tested, while the \openapi is used to list all the potentially testable elements. The metrics are computed as the ratio of the number of tested elements to
the total number of elements documented in the \openapi
specification.

%The main challenge we faced in the development of this module has been finding the relation between the two input sets. In fact, the expected structure could be the test set to be a subset of the specification set. After studying the problem and the tools under comparison, we noticed that the two set intersect but without any of the two containing the other one. 

%Relation between the test set and the specification set. The intersection represents the actually tested set, while the \emph{Not expected} represents both additional behaviours, not described in the \openapi specification, and properties tested by the object tools that were not in the specification. The \emph{Not tested} partition, instead, represents the described parts that have not been tested.

% In the simple scenario where the tested elements are contained in the specification elements, the computation of every coverage criterion is straightforward: it is the number of tested elements divided by the total elements in the specification. In a more common scenario, instead, it is important to distinguish the elements defined in the specification, that can consequently be counted for the coverage, from the elements tested by the tool but that are not part of the \openapi specification (\ie the discovered elements). 

% \michele{How do we compute coverage with discovered elements?}

Following the approach of Martin-Lopez et al.~\cite{MartinLopez2019TestCC}, metrics are divided into input and output metrics. Input metrics are computed based on the observed HTTP requests sent to the REST API under test, and comprise the following.
\begin{itemize}
\item {\bf Path coverage}: ratio of the number of tested paths to the total number of paths documented in the \openapi specification.

\item {\bf Operation coverage}: ratio of the number of tested operations to the total number of operations.

\item {\bf Parameter coverage}: ratio of the number of input parameters used by test cases to the total number of parameters.

\item {\bf Parameter value coverage}: ratio of the number of the exercised parameter values to the total number of possible values that parameters can assume. This metric applies, at the moment, only to domain-limited parameters such as \texttt{boolean} and \texttt{enum} types.

\item {\bf Request content-type coverage}: ratio of the number of tested content-types to the total number of accepted content-types. This metric is computed only when operations content-types have no wildcard (\eg \texttt{application/*}), because otherwise the number of accepted content-types would be unbounded.
\end{itemize}

Similarly, the following output metrics are computed based on the observed HTTP responses, emitted by the REST API under test.
\begin{itemize}
\item {\bf Status code class coverage}: a test suite reaches 100\% status code class coverage when it is able to trigger both correct status code (\ie the \twoxx class) and erroneous status codes (\ie \fourxx and \fivexx classes). %Conversely, if it only triggers status codes belonging to the same class (either correct or erroneous), the reached coverage is 50\%.

\item {\bf Status code coverage}: the ratio of the number of obtained status codes to the total number of status codes documented in the \openapi specification.

\item {\bf Response content-type coverage}: the ratio of the number of obtained content-types to the total number of response content-types documented in the \openapi specification. As for the input case, this metric is computed only when specific content-types are defined with no wildcard.
\end{itemize}

% {\bf Test coverage levels}
%Other than converge metrics, \rst also automatically computes Test Coverage Levels (TCLs), as defined in Martin-Lopez et al.~\cite{MartinLopez2019TestCC}. The TCL is a summary coverage indicator, ranking test suites based on the score they obtained for the defined metrics. In particular:
\rst also automatically computes Test Coverage Levels (TCLs), as defined in Martin-Lopez et al.~\cite{MartinLopez2019TestCC}. The TCL is a summary coverage indicator, ranking test suites based on the score they obtain for the defined metrics. TCLs are incremental: to achieve a TCL, the test suite also has to satisfy the requirements for the lower levels. In particular:

\begin{LaTeXdescription}
	\item[TCL0] is the base level, with no coverage requirements;
	\item[TCL1] requires the path coverage to be 100\%;
	\item[TCL2] requires the operation coverage to be 100\%;
	\item[TCL3] requires both request and response content-type coverages to be 100\%;
	\item[TCL4] requires parameters and status code classes coverages to be 100\%;
	\item[TCL5] requires the status code coverage to be 100\%;
	\item[TCL6] requires the body properties coverage to be 100\%;
	\item[TCL7] requires the operation flow coverage to be 100\%.
\end{LaTeXdescription}

\section{Running Example}
\label{sec:runningexample}

To better understand the use of \rst, we now present a running example involving a REST API and its (manually written) test suite, for which \rst will compute coverage metrics.

\subsection{The Pet Store and Its Test Suite}
The case study REST API is \textit{Pet Store}, a service to manage a collection of pets. The specification of the API describes the following 7 operations, arranged in 4 endpoints:

\begin{LaTeXdescription}
	\item[\texttt{POST} \!\! \texttt{/pet}] to enter data of a new pet;
	\item[\texttt{GET} \!\! \texttt{/pet}] to retrieve a list of all stored pets;
	\item[\texttt{PUT} \!\! \texttt{/pet}] to update data of a pet;
	\item[\texttt{GET} \!\! \texttt{/pet/findByStatus}] to search pets by status;
	\item[\texttt{GET} \!\! \texttt{/pet/findByTags}] to search pets by tag;
	\item[\texttt{GET} \!\! \texttt{/pet/\{id\}}] to retrieve a specific pet;
	\item[\texttt{DELETE} \!\! \texttt{/pet/\{id\}}] to delete all the data for a specific pet.
\end{LaTeXdescription}

We have manually prepared an example test suite for the Pet Store API consisting of five test cases based on the business logic of the API, one of which is reported in Listing~\ref{testcases}. They are five HTTP requests that can be executed, for instance, by means of Postman~\cite{Postman}.% \michele{Only one test is reported in the listing!}

The \openapi specification of the Pet Store and the dump of the HTTP requests/responses generated by the test suite are available in the \texttt{example} folder of the \rst repository.

\begin{figure}[t]
	\begin{lstlisting}[caption={A test case of our test suite.}, captionpos=b, language=http, numbers=none, linewidth=\linewidth, xleftmargin=5pt, label=testcases]
POST /v2/pet HTTP/1.1
Host: localhost:8080
Accept: application/json
Content-Type: application/json
Content-Length: 76
	
{
	"name": "doggie",
	"photoUrls": ["myphoto.com/doggie"]
}
	\end{lstlisting}
	%\vspace*{-10pt}
\end{figure}

The first two tests exercise the \texttt{GET} \!\! \texttt{/pet} operation. They obtain a successful response (\texttt{200} status code). The third and fourth tests exercise the \texttt{POST} \!\! \texttt{/pet} operation, and they obtain a \texttt{200} and \texttt{500} status code, respectively. The \texttt{500} internal server error is caused by a request (in Listing~\ref{testcases}) that misses the \texttt{category} mandatory parameter. This erroneous request is not properly handled by the Pet Store API, that throws an exception. Finally, a fifth request to the \texttt{PATCH} \!\! \texttt{/pet} operation obtains a \texttt{405} status code because such endpoint does not actually support the \texttt{PATCH} method: the test engineer would have wanted to use the \texttt{PUT} method instead, that operates similarly.

\subsection{Configuring and Launching \rst}

The five requests of our test suite, and the five related responses, have been recorded by the Burp proxy. Using the burp-dump plugin we exported the plain-text dumps to the \texttt{dumps} folder. Afterwards, we configured \rst with the following JSON configuration file (\texttt{config.json}):

\begin{lstlisting}[language=json, numbers=none, linewidth=\linewidth, xleftmargin=8pt, frame=l]
{
	"modules": "all",
	"specification": "/path/to/petstore.json",
	"dumpsDir": "/path/to/dumps",
	"reportsDir": "/path/to/reports",
	"dbPath": "/path/to/database.sqlite"
}
\end{lstlisting} % caption={The configuration of \rst.}, captionpos=b, 

We configured the tool with the modules to run (\texttt{modules}); where it can find the \openapi specification of the REST API under test (\texttt{specification}); where we have stored the dumps exported from Burp (\texttt{dumpsDir}\footnote{Absolute paths should be preferred for higher compatibility.}); where we want the output reports to be stored (\texttt{reportsDir}); and where to place the SQLite database (\texttt{dbPath}).

With the \texttt{modules} parameter set to \texttt{all}, we chose to run all the \rst modules. It is possible to run individually either the data collection module (with \texttt{dataCollection}), or the metrics computation module (with \texttt{statistics}).

Eventually, we have launched the tool from the folder \texttt{restats}, with the command: \ \texttt{python3} \!\! \texttt{app.py}.

\subsection{\rst Reports}

Once the execution of \rst is complete, its output is available in the chosen report folder as a collection of nine report JSON files. %containing the computed statistics, along with other interesting information to be analyzed by developers or test engineers.
The file \texttt{stats.json} summarizes the computed statistics for all the coverage metrics and the TCL of the test suite. The other eight files, named after the eight coverage metrics, describe in detail the output for each metric.

\begin{figure}[t]
	\begin{lstlisting}[caption={Part of \texttt{stats.json} regarding the operation coverage metric.}, captionpos=b, language=json, numbers=none, linewidth=\linewidth, xleftmargin=5pt, label=opecoverage]
{
	...
	"operationCoverage": {
		"raw": {
			"documented": 7,
			"documentedAndTested": 2,
			"totalTested": 3
		},
		"rate": 0.2857142857142857
	},
	...
	"TCL": 0
}
	\end{lstlisting}
	%\vspace*{-10pt}
\end{figure}

Listing~\ref{opecoverage} shows a fragment of \texttt{stats.json} with the operation coverage metric and the computed TCL. The remaining metrics are not shown for space reasons, however, they have the same structure.% \mariano{siccome c'e' spazio, completa la figura con le altre metriche. Se serve, questa figura potrebbe essere messa su 2 colonne} 
Observing Listing~\ref{opecoverage}, we can note that two operations, out of the seven documented in the specification, have been tested. Indeed, the computed \texttt{rate} is $0.2857$, that corresponds to $\frac{2}{7}$. In particular, the \texttt{documented} field reports the number of operations documented in the specification. The \texttt{documentedAndTested} field reports the number of operations that have been tested among those documented in the specification. Finally, the \texttt{totalTested} field reports the total number of tested operations, including those that are \emph{not} documented in the specification. This last field was included to reveal test cases that operate (knowingly or unknowingly) beyond what is documented.

%and the \texttt{tested} fields are both set to $2$. , that in this case corresponds to the \texttt{tested} field, actually describes the number of operation 

Listing~\ref{rstreport} shows the content of one of the eight detailed reports (specifically, the operation coverage metric report) that \rst emits for each coverage metric. The \texttt{documentedAndTested} section  reports what operations have been tested, among those documented in the specification. The \texttt{documentedAndNotTested} section reports what operations, among those documented in the specification, have \textit{not} been tested. Finally, undocumented operations that were tested, if any, will be reported in the \texttt{notDocumentedAndTested} section.

\subsection{How Are These Reports Useful?}

The reports computed by \rst come in handy in a number of ways. By looking at Listing~\ref{opecoverage}, we could easily identify that not all the available operations in the REST API have been properly tested. Only two operations  out of seven have been tested, which prevents the test suite to be considered complete. This should motivate a developer to write more test cases to test, at least once, all the documented operations. 

Moreover, the \texttt{notDocumentedAndTested} section in Listing~\ref{rstreport} indicates that the developer inappropriately used the \texttt{PATCH} method with the \texttt{/pet} endpoint in a test case.

\begin{figure}[t]
	\begin{lstlisting}[captionpos=b, language=json, numbers=none, linewidth=\linewidth, xleftmargin=5pt, caption={The detailed output for the operation coverage metric.}, label=rstreport]
{
	"documentedAndTested": {
		"/pet": ["get", "post"]
	},
	"documentedAndNotTested": {
		"/pet": ["put"],
		"/pet/findByStatus": ["get"],
		"/pet/findByTags": ["get"],
		"/pet/{petId}": ["get", "delete"]
	},
	"notDocumentedAndTested": {
		"/pet": ["patch"]
	}
}
	\end{lstlisting}
	%\vspace*{-10pt}
\end{figure}
\section{Related Work}
\label{sec:related}

In literature, several tools have been presented to measure test coverage for specific domains, like Java programs (the JaCoCo library~\cite{JaCoCo}) or Android apps (ACVTool~\cite{ACVtool2018} and COSMO~\cite{COSMO2021}). However, to the best of our knowledge, no tool has been presented for assessing the coverage of REST APIs test cases. This might be due to the fact that, often, coverage tools rely on \emph{code coverage}, which is not always applicable in the context of REST APIs. Indeed, API source code analysis might be complex, e.g., when a micro-service architecture includes many dynamically allocated components possibly from different vendors; or not feasible, e.g., in the case of closed-source APIs.
The only work dealing with test coverage of REST APIs is the theoretical framework of Martin-Lopez et al.~\cite{MartinLopez2019TestCC}, (presented as background in Section~\ref{sec:architecture}), but it does not come with a usable implementation. 

Concerning REST API testing, we can find semi-automatic tools (like REST-Assured~\cite{RESTAssured}, SoapUI~\cite{Soap}, Postman~\cite{Postman}), but they do not provide a measure for assessing test case coverage. Similarly, automatic test case generation tools for REST APIs (like QuickREST~\cite{Karlsson2020QuickREST}, RESTler~\cite{Atlidakis2019RESTlerSR} RestTestGen~\cite{Viglianisi2020RESTTESTGENAB}), report what operations could be tested and what status code have been observed in the responses, but in a tool-specific way that does not follow a common formal definition of test coverage.

\section{Conclusion}
\label{sec:conclusion}

In conclusion, the present paper introduce \rst, a tool that implements the test coverage metrics originally proposed by Martin-Lopez et al.~\cite{MartinLopez2019TestCC}. \rst computes the coverage metrics of the REST API under test adopting a black-box perspective, namely relying only on the OpenAPI specification of the REST API under test. In particular, \rst computes the coverage metrics based on the network log that contains the plain-text HTTP requests and responses exchanged when running test cases. This makes our tool easily adaptable to multiple contexts, supporting services written in any programming language, and any testing technology and tool. 

\rst can be used to measure test case coverage in multiple adoption scenarios, either by programmers at development time or by stakeholders before deployment in production. Finally, \rst can help researchers in assessing automated test case generation techniques, highlighting strengths and weaknesses of the adopted generative approach. Indeed, \rst has been used in~\cite{scam2021} to compare four state-of-the-art black-box REST APIs testing tools.

\bibliographystyle{IEEEtran}
\bibliography{bib}

\end{document}